# Flexible Control Strategy of DC Bus for AC-DC Hybrid Microgrid with Electric Vehicle

Wang Jingsong
DERI Energy Research Institute
Nanjing, China
e-mail: wangjingsong@deri.energy

Gong Chengming
DERI Energy Research Institute
Nanjing, China
e-mail: gongchm@live.cn

*Abstract*—As a new type of microgrid structure, AC-DC hybrid microgrid can efficiently consume new energy distributed generator based on photovoltaics, which is very suitable for microgrid systems with electric vehicles as the main load. Unlike the AC microgrid, the DC bus of the AC-DC hybrid microgrid is a low-inertia system. How to improve the DC bus inertia of the AC-DC hybrid microgrid system and the stability of the DC bus voltage become particularly important. Based on this, this paper presents a method for flexible control strategy of microgrid bus voltage based on multi-node droop. By considering the P / U droop characteristics of DC ports of power electronic equipment such as energy storage and electric vehicle charging-discharging equipment, different types of distributed generator are comprehensively considered. The power reserve rate and energy reserve rate, through curve shift and other adjustment methods, improve the DC bus inertia, which effectively guarantees the stability of the microgrid system voltage. The validity of the proposed method is verified by building a matlab / simulik simulation system.

*Index Terms*-- Electric vehicles; AC-DC hybrid microgrid; DC bus; Stability control

## I. Introduction

In May 2018, Caixin reported that global warming could lead to the complete disappearance of arctic sea ice in summer within 22 years, and the development of electric vehicles is an inevitable trend [1]. In 2015, China's National Development and Reform Commission issued the "Guide to the Development of Electric Vehicle Charging Infrastructure (2015-2020)", which aims to promote electric vehicles in China. However, the primary energy is mainly fossil energy, and it is difficult to reveal the advantages of energy-saving and emission reduction for electric vehicles directly connected to the grid [2]-[3].

To achieve real energy-saving and emission reduction, we build Photovoltaic (abbreviated as PV) and energy storage in electric vehicle charging stations to form an AC-DC hybrid microgrid[4]. We choose DC bus to reduce the number of converters and energy loss during energy interaction[5], which can achieve the most efficient consumption of new energy generation.

The stability of the DC bus is the basis for the stable operation of the AC-DC hybrid microgrid [6]-[8]. In the field of microgrid control, droop control has become a mainstream control strategy. The AC droop control method has been widely used, and some engineering pilot applications have been applied [9]-[11]. Some scholars have also done some research on DC droop control methods [12] -[14]. But research and engineering validation are still needed about how electric vehicles effectively participating in the stable control of the DC bus voltage, at the same time, comprehensively considering of the characteristics of electric vehicles, energy storage, and PV.

Based on the research of the existing DC bus droop control method, this paper proposes an engineering applied method of flexible control strategy of DC bus voltage stability in which electric vehicles directly participate in. And this paper verifies the correctness of the method by minimum system test.

## II. AC-DC Hybrid Microgrid System with Electric Vehicles

The structure of AC-DC hybrid microgrid system with electric vehicles is shown in Figure 1. The primary equipment of the microgrid includes PV, energy storage, electric vehicle charging-discharging equipment, AC/DC bidirectional power converter (abbreviated as AC/DC), power distribution switch and loads. The main secondary equipment of microgrid includes a microgrid controller and microgrid energy cloud.

*A. Primary system structure*

As we can see in Figure 1, PV, energy storage and the electric vehicle charging-discharging equipment exchange energy through the DC bus. The DC bus and the AC bus interact through the AC/DC. The system is connected to the large power grid through Point of Common Coupling (abbreviated as PCC), and also has some AC loads on the AC side.

*B. Secondary system structure*

The secondary system consists of two parts: a microgrid controller and microgrid energy cloud.

The microgrid controller acquires the DC bus voltage through a voltage Hall sensor, and collects PV, energy storage, AC/DC, and electric vehicle charging-discharging equipment output currents through current Hall sensors. Through CAN protocol the microgrid controller communicates with the photovoltaic optimizer, energy storage converter, charging-discharging module and AC/DC, acquiring the operating parameters of each device and distributing control instructions.

The microgrid controller communicates with the microgrid energy cloud through Internet.

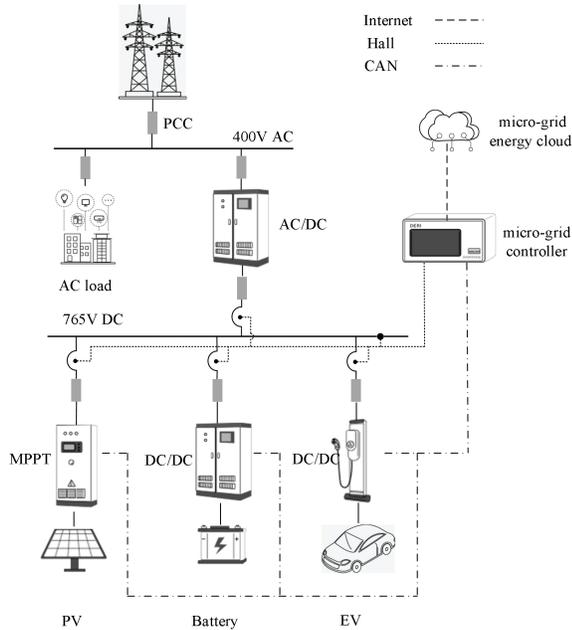

Figure 1. AC-DC hybrid microgrid system structure

## III. FLEXIBLE CONTROL STRATEGY OF DC BUS

The DC bus flexible control strategy of the AC-DC hybrid microgrid is completed by the power electronic equipment and the microgrid controller connected to the DC bus. The detailed description is as follows.

### A. Power Electronic Equipment Control

The power electronic equipment connected to the DC bus can be divided into two types: AC/DC and DC/DC. Between them, AC/DC is used to realize bidirectional power interaction with the AC system, and DC/DC is used to realize energy interaction with energy storage, electric vehicles and photovoltaic power generation equipment.

#### 1) DC/DC

The topology and control loop of DC / DC is as follows:

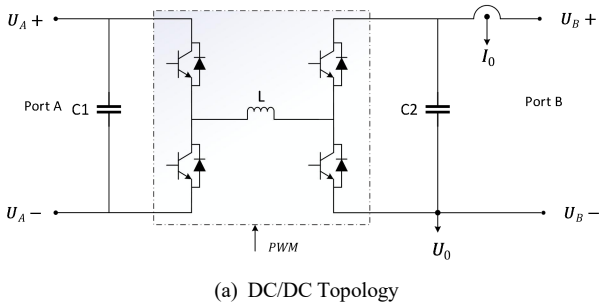

(a) DC/DC Topology

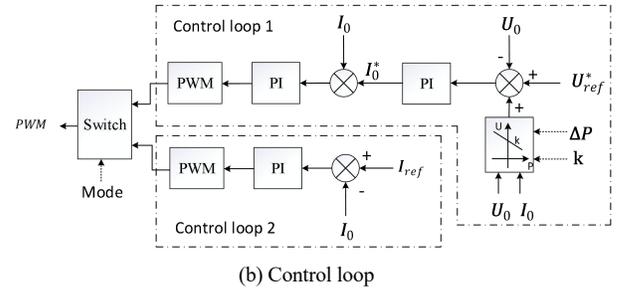

(b) Control loop

Figure 2. The topology and control loop of DC / DC

The DC/DC applied in this paper uses a two-port four-switch Buck-Boost topology, which can achieve bidirectional buck-boost conversion. Port A is connected to electric vehicles, energy storage and other equipment, and port B is connected to the DC bus. Considering that this paper only studies the control of the DC bus, the control logic of port B is given, as shown in Figure 2 (b).

It can be seen that the control logic of port B is composed of two control logics, and the control output selection of different control loops is realized through a switch. Among them, control loop 1 guarantees the stability of the output voltage by means of voltage outer loop and current inner loop. A drooping link is added outside control loop 1 to ensure that the port voltage changes linearly with the output power. The control logic of the droop loop will be described in detail in the following sections. Control loop 2 is a current loop to ensure that the output current is output in accordance with the set current. The current loop cannot support the bus voltage.

#### 2) AC/DC

The AC / DC topology and control loop are as follows:

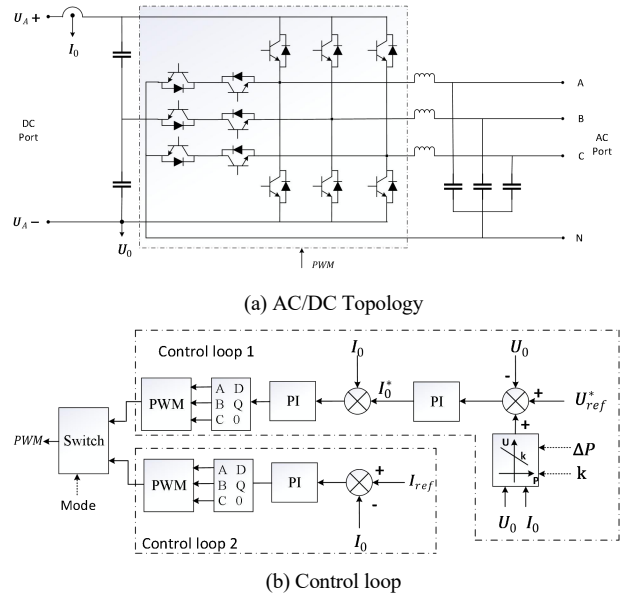

(a) AC/DC Topology

(b) Control loop

Figure 3. The topology and control loop of DC / DC

AC/DC equipment adopts T-type three-level topology, which can realize bidirectional power interaction. 3 (b) gives the control logic of DC port, which is similar to DC/DC and won't be detailed here.

3) Droop control

The expression of the droop control loop is as follows:

$$U_0^* = U_{ref}^* - k(P_{out} - \Delta P) \quad (1)$$

In the formula, $U_0^*$ is the reference voltage of the voltage outer loop. $U_{ref}^*$ is the rated voltage set by the power electronic device. k is the droop coefficient. $P_{out}$ is the current output power, and $\Delta P$ is the power change rate adjusted by the microgrid controller.

It can be seen that when $\Delta P$ is set to 0, the actual reference voltage $U_0^*$ of the voltage outer loop will decrease as the output current increases, which is the droop control characteristics. When $\Delta P$ is not 0, not only the actual reference voltage $U_0^*$ of the voltage outer loop decreases as the output current increases, but also the rated voltage $U_{ref}^*$ set by the power electronic device changes. That is, the droop curve has shifted.

## B. Flexible Control Strategy of DC Bus Voltage

The flexible control strategy of the bus voltage is realized by the microgrid controller, and its control logic diagram is shown in Figure 4.

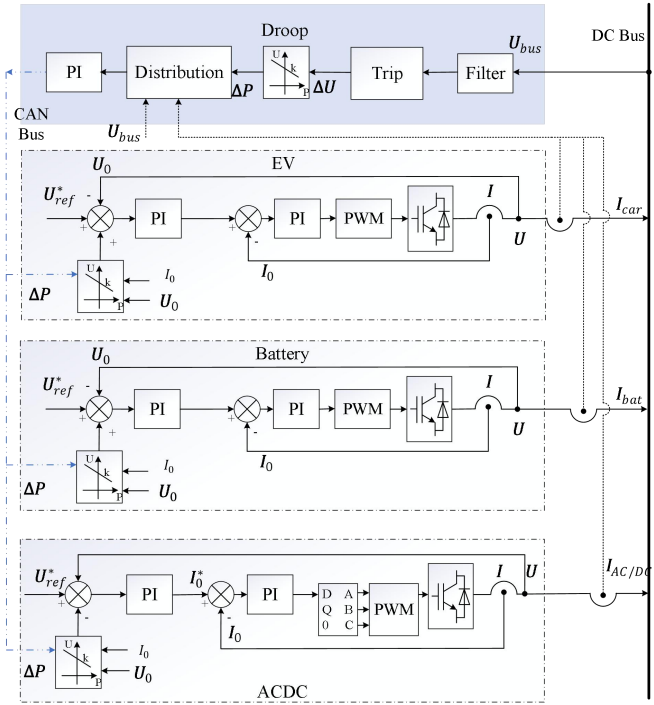

Figure 4. Flexible control logic block diagram

The core idea of flexible control is to calculate the shortfall value of DC bus system power through the change of DC bus voltage, and then select the appropriate power electronic equipment to compensate for the change in power, and finally achieve the purpose of DC bus voltage stabilization.

Figure 4 shows a three-node system bus flexible control logic block diagram. It can be seen that the control method can be decomposed into three parts: start-up logic, system droop curve calculation logic, and power allocation logic. Separate explanation is as follows.

1) Trip

The bus voltage start logic is responsible for triggering the process and outputting the current voltage difference. Take the bus voltage as U, set the voltage deviation set-value to $\Delta U_{set}$. The setting principle is that the equipment connected to the DC bus can operate normally within this range. The bus rated voltage is $U_n$. The voltage start criterion can be described as:

$$|U - U_n| > \Delta U_{set} \quad (2)$$

At this time, the output voltage fluctuation value is $\Delta U = U - U_n$;

2) Droop

The main function of the droop curve calculation logic is to calculate the droop curve of the microgrid system in real time, and output the unbalanced power value of the system according to the voltage value output by the start logic.

Take the control system of FIG. 4 as an example, the control system is composed of two DC/DC and one AC/DC in parallel. Each power electronic device has a U/P droop due to its virtual impedance, so the virtual impedance of the DC bus system is the parallel connection of the impedance of the three power electronic modules.

At the same time, it should be noted that the virtual impedance of the device can be calculated into the system impedance only when it is running in voltage source mode, and its virtual impedance cannot be calculated in parallel as the impedance of the DC bus system when it is running in current source mode or in a fault state.

3) Distribution

The droop curve calculation logic has calculated the system power fluctuation value, and the power allocation logic allows the user to achieve the final power allocation. The calculation method is as follows:

a) Each drooping node can be artificially set a weight coefficient $\delta$ ($\delta \in [0,1]$). Increase the setting if you want the node to participate in the adjustment first;

b) The device is in a locked state, that is, if the node state is abnormal and the regulated power cannot be output, then the device is locked and will not participate in adjustment;

c) Calculate the power reserve rate. take $P_n$ as the rated output power of the power electronic equipment participating in the control, and $P$ as the current output power, then the power reserve ratio of the droop device at this moment is:

$$\beta = |P_n - P| / P_n \quad (3)$$

d) Calculate the energy reserve rate. For the energy storage device, the energy reserve rate is calculated as follows:

$$\gamma = (W_n - W) / W_n \quad (4)$$

Where $W_n$ is the rated capacity of energy storage. W is the current capacity value of energy storage. For the actual control system, the energy reserve rate can be directly replaced by the stored energy SOC (State-of-charge) value ([0,100]). For non-energy storage power electronic equipment set γ=100;

e) For a non-locked device, the competition coefficient is:

$$\partial = \delta \times \beta \times \gamma \quad (5)$$

For the locked device, $\partial=0$;

f) Sort the competition coefficients of each drooping node in descending order;

g) The total power fluctuation value calculated by the droop curve fitting block is subjected to power distribution according to the competition coefficient calculated in step e). The principle of allocation is to prioritize the allocation of power to one or more drooping nodes that are ranked first. And the microgrid controller passes allocation values through the CAN bus to each node to complete this control.

## IV. TEST VERIFICATION

### A. Test system

In order to verify the effectiveness of the method described in this article, a test system was built based on Matlab / Simulink to verify the method described in this article. The schematic diagram of the system is as follows:

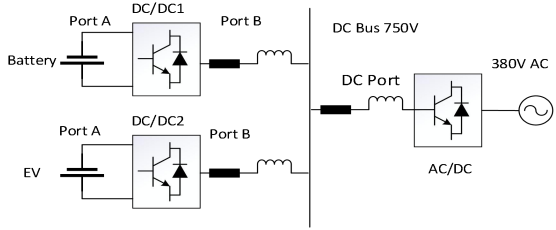

Figure 5. Test system structure

The initial equipment parameters and control parameters in the minimum system are shown in the table.

TABLE I. SYSTEM PARAMETERS

| device | $U_n(V)$ | $P_n(kW)$ | SOC | $k(kW/V)$ |
|---|---|---|---|---|
| AC/DC | 750 | 60 | - | 1 |
| BATTERY | 750 | 15 | 50 | 4 |
| EV | 750 | 15 | 50 | 4 |

In the table: $U_n$ is the rated voltage, $P_n$ is the rated capacity of the micro-power supply, and SOC is the power of each micro-power supply. For systems without energy storage, the default SOC is 1, k is the droop coefficient of the micro-power supply. The voltage start stable range is [748, 752] V.

### B. Test Case

1) *Case 1: EV charging under grid connection*

The initial test conditions are: DC / DC1 stabilizes the B port voltage, DC / DC2 works in the current source mode, and AC / DC works in the stable DC port voltage mode. After the test started, DC/DC2 charges the electric vehicle with a constant power of 15kW. The test results are as follows.

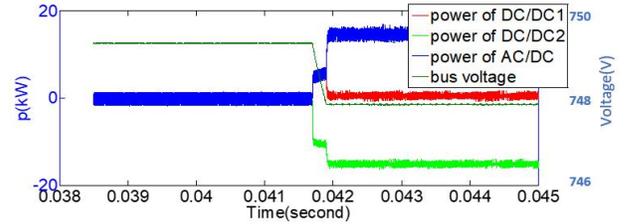
(a) Voltage and power adjustment process

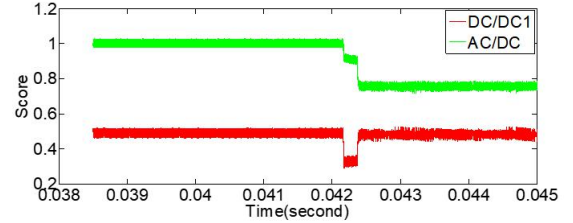
(b) competition coefficient

Figure 6. Test results of case 1

It can be seen that when DC/DC2 is charged at a constant power of 15kW, the system bus voltage starts to drop due to the drooping effects of DC/DC1 and AC/DC ports. When the voltage drops below 748V, the flexible control strategy of the bus voltage comes into play. As can be seen from the competition scores of DC/DC1 and AC/DC in Figure (b), AC/DC always has a high competition coefficient. The system adds AC/DC power output to supplement the 15kW charging power shortage. DC/DC1 does not output power under steady state.

2) *Case 2: Charge adjustment process in off-grid state*

The initial test conditions are: DC/DC1 stabilizes the B port voltage, DC/DC2 works in the current source mode, and AC/DC works in the current source mode. At this time, the DC bus voltage is maintained stable by DC/DC1. After the test started, DC/DC2 charges the electric vehicle with a constant power of 15kW. The test results are as follows.

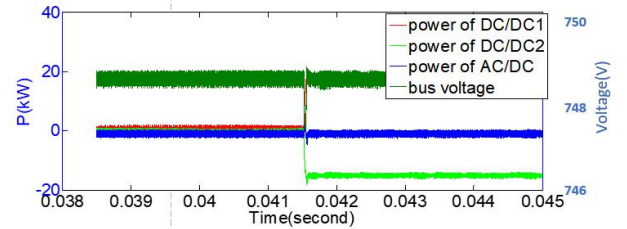

Figure 7. Test results of case 2

Because the system has only one voltage source, when DC/DC2 is charged at a constant power of 15kW, the DC bus voltage drops due to the droop of DC/DC1. When the voltage drops below 748V, DC/DC1 increases the power output by 15kW, and the system bus returns to [748, 752] V, the system enters steady state operation.

3) *Case 3: Support AC load in off-grid state*

The initial test conditions are: DC/DC1 and DC/DC2 stabilize the voltage on port B at the same time, and AC/DC

works in current source mode. After the test started, AC/DC generated power to the AC grid with a constant power of 15kW. The test results are as follows.

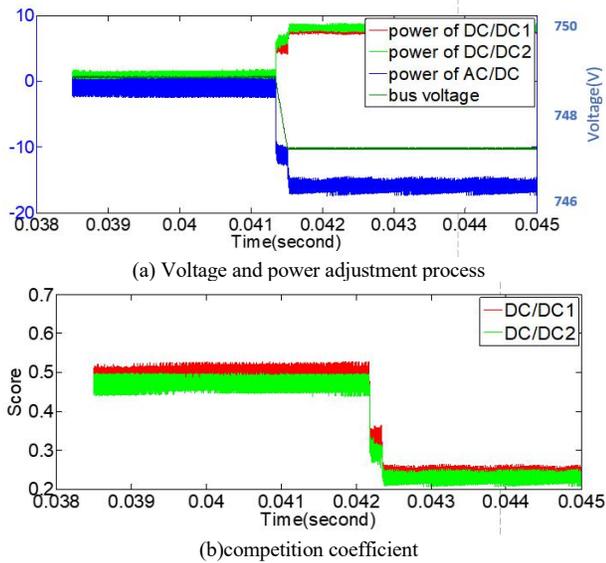

(a) Voltage and power adjustment process

(b) competition coefficient

Figure 8. Test results of case 3

Similar to the previous two test conditions, the DC bus voltage began to drop due to the droop effect of the DC/DC1 and DC/DC2 ports at the beginning stage. The flexible control strategy of the bus voltage was started after the start value was lowered, due to the original DC/DC1 and DC/DC2 The setting parameters are the same (you can also see that the competition coefficients of the two devices are basically the same from Figure (b)). The average output of DC/DC1 and DC/DC2 is about 7.5kW. The system enters a steady state and the bus voltage is stable at the setting interval.

From the three experimental results, it can be seen that, compared with power electronic devices without droop loops, when power sudden changes occur in the system, each power electronic device can automatically distribute power according to their respective virtual impedance, and jointly support the stability of the bus, and the flexible control strategy of the bus voltage in different operating modes such as grid-connected mode and off-grid mode, power electronic devices in different operating modes can be coordinated, and devices with strong adjustment capabilities are preferred to ensure that the system voltage is within the set range, which greatly improves the inertia of the DC bus and the power quality of the DC bus.

V. CONCLUSION

This article focuses on the stability of the DC bus voltage of an AC-DC hybrid microgrid system with electric vehicles. It makes full use of the P/U droop characteristics of DC ports of power electronic equipment such as energy storage and electric vehicle charging-discharging equipment, and considers different types of new energy distributed generators' power reserve ratio and energy reserve ratio. Through flexible adjustment methods such as curve shift, we improve the DC bus inertia, effectively guarantee the stability of the microgrid system voltage, and improve the power quality of the DC system. The method described in this article has been widely used in the TELD company's integrated photovoltaic-energy storage-charging system, which provides a strong guarantee for the safe and stable operation of the system and brings good economic benefits.